\documentclass[aps]{revtex4}%
\usepackage{amsfonts}
\usepackage{amsmath}
\usepackage{amssymb}
\usepackage{graphicx}%
\providecommand{\U}[1]{\protect\rule{.1in}{.1in}}

\begin{document}
\preprint{ }
\title[Short title for running header]{Brane cosmology with variable tension}
\author{Jos\'{e} Antonio Belinch\'{o}n}
\affiliation{Dpt. Matem\'{a}ticas. Universidad de Atacama. Av. Copayapu 485, Copiap\'{o},
Chile }
\author{Sami Dib}
\affiliation{Max Planck Institute for Astronomy, K\"{o}nigstuhl 17, 69117, Heidelberg,
Germany }
\pacs{PACS number}

\begin{abstract}
We study brane-world models and demonstrate that such models do not admit
self-similar solutions through the matter collineation approach. By
introducing the hypothesis of variable brane tension, $\lambda,$ we outline
the new effective field equation (EFE) in the most simple case (symmetric
embedding) under the assumption that the fundamental constants in 5D are
constants. In this case, we find the exact form that each physical quantity
may take in order that the EFE become invariant under scale transformations.
By taking into account such assumptions, we find that in 4D, the gravitational
constant $\kappa^{2}\thicksim\lambda$ while the cosmological constant
$\Lambda\thicksim\lambda^{2}$ are always decreasing. These results are quite
general and valid for any homogeneous self-similar metric. Nevetheless, the
study of the EFE under scale symmetries suggests that $\rho\thicksim\lambda$
(as a functional relationship). This allows to get a growing $\kappa^{2}$ but,
in this case, the fundamental constants in 5D must vary as well. We outline a
toy model allowing such a possibility.

\end{abstract}
\date{\today}
\maketitle

\section{Introduction}

Recent cosmological observations of Supernovae Type Ia (SNIa) at high redshift
show evidence of an accelerated expansion of the Universe at late times (see
for instance \cite{SNIa1}-\cite{SNIa3}). This is also supported by the
observations of anisotropies in the CMB as well as in baryon accoustic
oscillations (BAO).

Several ideas have been proposed in order to explain such acceleration. An
interesting paradigm is to consider extra dimensions of space-time which could
be the source of the current accelerated expansion. For instance, the
Dvali-Gabadadze-Porrati model \cite{DGP1,DGP2} generates a natural accelerated
expansion with a geometrical threshold associated to a five-dimensional
space-time. Another example are the models proposed by Randall and Sundrum
\cite{RS1,RS2} (hereafter RS models). They imagined our four-dimensional (4D)
space-time could be a three-brane embedded in a 5-dimensional (5D) space-time
(the bulk). According to the brane-world scenario, the physical fields
(electromagnetic, Yang-Mills etc.) observed in our 4D Universe are confined to
the three-brane. Only gravity can freely propagate both on the brane and in
the bulk, with the gravitational self-couplings not significantly modified.
Even with the fifth dimension uncompactified, standard 4D gravity can be
reproduced on the brane in a certain limit \cite{RM2010}. At very high
energies, in the presence of large 5D curvatures, significant deviations from
the standard Einstein theory could occur in brane-world models, due to the
nonstandard model 5D fields, and possible asymmetric embeddings of the brane
into the bulk. Nevertheless, RS models drive a late time acceleration only by
adding a dark energy field \cite{maga1}. In order to alleviate this
discrepancy between the theoretical predictions and the observations, it has
been suggested that the brane tension could be variable. In this way, in
\cite{maga2} the authors investigate a brane model with variable brane tension
as a function of redshift called chrono-brane. They showed that the model is
consistent with the observational data and that it can drive the Universe to
an accelerated phase at late times.

Brane-worlds with non-constant tension have been proposed by several authors
(see for instance \cite{NCD1}-\cite{Gerg2}). This new degree of freedom allows
for evolving gravitational and cosmological constants, the latter being a
natural candidate for dark energy. For cosmological branes, a variable brane
tension leads to several important consequences that have been studied for
example in (\cite{Gerg2} and \cite{CC1,CC2,CC3}). But this is not the only way
for introducing the hypothesis of evolving constants in 4D. Higher-dimensional
theories imply that some constants, such as the gravitational constant and the
strength of the gauge-couplings, are not fundamental constants. Instead, they
are related to the sizes of the extra-dimensional space, which are moduli
fields in the four-dimensional effective theory \cite{Brax2003}. In another
approach, Ponce de Leon \cite{PdL1,PdL3} study the possibility that the
variation of the fundamental constants, in the framework of brane-world
models, are provoked by the fifth dimension described by a scalar function
varying with time. Other works contemplate the variation of the fundamental
constant, as well as other constants such as the speed of light \cite{Youm1}%
-\cite{BKM}.

As we have discussed above, the brane-world models with variable tension fit
better with the observations than those with constant tension. Therefore our
goal is to study brane-world cosmological models from different points of
view. In the first place we determine that the models with constant tension
cannot admit self-similar solutions. Secondly, we study whether the models
with variable tension admit such kind of solutions. If the field equations are
invariant under scale transformations then study how must vary each physical
quantity, putting special emphasis in determining how the gravitational and
the cosmological \textquotedblleft constants\textquotedblright\ vary. Thirdly,
we explore other frameworks that allow the fundamental constants to vary in 5D.

The present paper is organized as follows. In Section II, we introduce the
basic effective field equations (EFE) of the brane-world cosmological models.
Then, we determine if the outlined EFE are invariant under scaling
transformations. In this way, we show that the model does not admit homothetic
solutions. The basic idea behind this geometrical procedure is a
generalization of the Dimensional Analysis (DA). In Section III, we introduce
the hypothesis of a variable brane tension. We start by formulating the new
EFE in the most basic case and under the assumption that the fundamental
constants behave as true constants in 5D. The hypothesis of a time varying
brane tension implies that the gravitational and the cosmological constants
must vary with time. Then, we determine the exact form that each physical
quantity may take in order for the new EFE to admit exact self-similar
solutions through the matter collineation approach, that is, we study if the
new EFE are invariant under scaling transformations. This simple approach
allows us to find exact solutions for a wide variety of homogeneous
cosmological models since the results are quite general and valid for any
self-similar metric. As examples, we study two cases by finding exact
solutions to these particular models. In section IV, in a phenomenological
way, we outline the EFE of a toy model that allows the fundamental constants
in 5D to vary. In this way, we are able to generate new cosmological
scenarios. In Section V, we discuss our findings and conclude.

\section{Field equations in the brane-world models}

In a five dimensional space-time (called the bulk), it is possible to confine
matter fields in a four dimensional hypersurface (called 3-brane). The
effective Einstein equations on the bulk are \cite{sms2},\cite{sms1}%
,\cite{Maartens1} and \cite{RM2010}:%
\begin{equation}
G_{ab}^{5}=\kappa_{5}^{2}T_{ab}^{5}-\Lambda_{5}g_{ab}^{5},\label{FE1}%
\end{equation}
where $\Lambda_{5}$ is the 5D cosmological constant which is assumed to be
negative and $T_{ab}^{5}$ is the total energy-momentum tensor on the brane
(vacuum and matter) defined by%
\begin{equation}
T_{ab}^{5}=\delta\left(  \chi\right)  \left(  T_{ab}^{m}-\lambda
g_{ab}\right)  ,\label{DT}%
\end{equation}
where $T_{ab}^{m}$ represents ordinary matter and $\lambda$ is the vacuum
energy density interpreted as the tension of the brane, considered as
\textquotedblleft constant\textquotedblright, with the induced metric
$g_{ab}.$ The Dirac delta function indicates the fact that the matter is
confined in the hypersurface $\chi=0.$

The basic equations on the brane are obtained by projections onto the brane
with Gauss-Codazzi equations, Israel junction condition and assuming a $Z_{2}$
symmetry of the bulk\textbf{.} Then the effective Einstein equations (4D) on
the brane are:%
\begin{equation}
G_{ij}=\kappa^{2}T_{ij}^{m}+\kappa_{5}^{4}S_{_{ij}}-\mathcal{E}_{_{ij}%
}\,-\Lambda g_{_{ij}}, \label{FE4D}%
\end{equation}
where $\kappa^{2}$ is the 4D gravitational coupling \textquotedblleft
constant\textquotedblright\ given by $6\kappa^{2}=\lambda\kappa_{5}^{4}.$ For
attracting gravity, the brane tension $\lambda$ should be positive,
$\lambda>0$. There are several constraints which have been obtained for the
brane tension (see for instance \cite{Ten1,Ten2}). The cosmological constant
$\Lambda,$ is defined by $\Lambda=\kappa_{5}^{2}\left(  \Lambda_{5}+\kappa
_{5}^{2}\lambda^{2}/6\right)  /2.$

The above effective field equations (\ref{FE4D}) have two new correction
terms. The first correction term is the quadratic term $S_{_{ij}}$ arising
from the extrinsic curvature term in the projected Einstein tensor, and is
given by%
\begin{equation}
S_{_{ij}}={{\frac{1}{12}}}T_{\alpha}{}^{\alpha}T_{_{ij}}-{{\frac{1}{4}}%
}T_{i\alpha}T^{\alpha}{}_{j}+{{\frac{1}{24}}}g_{_{ij}}\left[  3T_{\alpha\beta
}T^{\alpha\beta}-\left(  T_{\alpha}{}^{\alpha}\right)  ^{2}\right]  .\label{S}%
\end{equation}
The second correction term, $\mathcal{E}_{_{ij}},$ is the projection of the 5D
Weyl tensor $C_{abcd}$, onto the brane, and is defined as%
\begin{equation}
\mathcal{E}_{_{ij}}=C_{ijcd}n^{c}n^{d},\label{U}%
\end{equation}
and encompasses nonlocal effects from the free gravitational field in the
bulk. This term vanishes if the bulk spacetime is purely anti-de Sitter. The
only general known property of this nonlocal term is that it is traceless.
$n^{d}$ is the unit normal to the hypersurface $\chi=0.$

The Einstein equation in the bulk, $\nabla^{b}T_{ab}^{5}=0,$ also implies the
conservation of the energy momentum tensor of the matter on the brane, that
is, $\nabla^{j}\left(  \kappa^{2}T_{ij}^{m}\right)  =0.$ In this way, we have
a new constraint on $S_{_{ij}}$ and $\mathcal{E}_{_{ij}}$ as%
\begin{equation}
\kappa_{5}^{4}\nabla^{j}\left(  S_{_{ij}}\right)  =\nabla^{j}\left(
\mathcal{E}_{_{ij}}\,\right)  \label{con2}%
\end{equation}

The symmetry properties of $\mathcal{E}_{ij}$ (the bulk Weyl tensor) imply
that, in general, we can decompose it irreducibly with respect to the
4-velocity $u^{i},$ in the following way \cite{Maartens1,RM2010}:
\begin{equation}
\mathcal{E}_{ij}={\frac{-6}{\kappa^{2}\lambda}}\left[  \mathcal{U}\left(
u_{i}u_{j}+{{\frac{1}{3}}}h_{ij}\right)  +\mathcal{P}_{ij}+\mathcal{Q}%
_{i}u_{j}+\mathcal{Q}_{j}u_{i}\right]  ,\label{DefU}%
\end{equation}
where the constant, $-6/{(\kappa^{2}\lambda),}$ is introduced ad hoc for
dimensional considerations. $\mathcal{U}$ is a scalar quantity, called dark
energy density, $\mathcal{P}_{ij}$ is a spatial symmetric and trace-free
tensor and $\mathcal{Q}_{i}$ is a spatial vector. $h_{ij}$ is the projection
tensor orthogonal to $u^{i}$ on the brane. By using the constraint equation
(\ref{con2}), it is possible to derive the evolution equations for
$\mathcal{U}$ and $\mathcal{Q}_{i}$ but not for $\mathcal{P}_{ij},$ so the
system of equations on the brane is not in general closed. In this work, for
simplicity, we only consider a flat FLRW metric, that is, homogeneous and
isotropic, so $\mathcal{Q}_{i}=\mathcal{P}_{ij}=0,$ and therefore, the only
non-zero contribution from the 5D Weyl tensor from the bulk is given by the
scalar quantity $\mathcal{U}.$ We would like to point out that other authors
(see for instance \cite{Gerg1}) use other definitions for $\mathcal{E}_{ij}$
such as%
\begin{equation}
\mathcal{E}_{ij}=-\kappa^{2}\mathcal{U}\left(  u_{i}u_{j}+{{\frac{1}{3}}%
}h_{ij}\right)  ,\label{GerU}%
\end{equation}
the unique difference are the coupling constants, but they change the nature
of $\mathcal{U}$ and its physical dimensions and therefore the cosmological
implications as we will show below.

Therefore the EFE on the brane are defined by:%
\begin{align}
G_{ij}  &  =\kappa^{2}T_{ij}^{m}+\kappa_{5}^{4}S_{_{ij}}-\mathcal{E}_{_{ij}%
}\,-\Lambda g_{_{ij}},\label{FE4D1}\\
\nabla^{j}T_{ij}^{m}  &  =0,\label{FE4D2}\\
\kappa_{5}^{4}\nabla^{j}\left(  S_{_{ij}}\right)   &  =\nabla^{j}\left(
\mathcal{E}_{_{ij}}\,\right)  . \label{FE34D}%
\end{align}

We assume a flat FLRW metric given by
\begin{equation}
ds^{2}=-dt^{2}+a(t)^{2}(dx^{2}+dy^{2}+dz^{2}),\label{FLRW}%
\end{equation}
and a perfect fluid whose energy-momentum tensor on the brane is defined by%
\begin{equation}
T_{ij}^{m}=\left(  \rho+p\right)  u_{i}u_{j}+pg_{ij},
\end{equation}
where $\rho$ is the energy density and $p$ is the isotropic pressure, with the
equation of state $p=\gamma\rho,$ with $\gamma$  being a constant$.$ Under all
these assumptions, the effective gravitational field equation EFE
(\ref{FE4D1}-\ref{FE34D}), with $\mathcal{E}_{ij}$ defined by Eq.
(\ref{DefU}), become%
\begin{align}
3H^{2} &  =\kappa^{2}\rho+\frac{\kappa^{2}}{2\lambda}\rho^{2}+\frac{6}%
{\lambda\kappa^{2}}\mathcal{U}+\Lambda,\label{FRW1}\\
2\dot{H}+3H^{2} &  =-\kappa^{2}p-\frac{\kappa^{2}}{2\lambda}\left(  2\rho
p+\rho^{2}\right)  -\frac{2}{\lambda\kappa^{2}}\mathcal{U}+\Lambda
,\label{FRW2}\\
\dot{\rho}+3\left(  \rho+p\right)  H &  =0,\qquad\rho=\rho_{0}a^{-3\left(
\gamma+1\right)  },\label{FRW3}\\
\mathcal{\dot{U}}+4H\mathcal{U} &  =0,\qquad\mathcal{U=U}_{0}a^{-4}%
,\label{FRW4}%
\end{align}
where $\cdot=d/dt,$ $H=\dot{a}/a.$ Eq. (\ref{FRW4}) is obtained from Eq.
(\ref{con2}), since for a perfect fluid, the left hand side of Eq.
(\ref{con2}) vanishes due to the energy-momentum conservation of the fluid.
Hence, in this case the non-local energy density obeys the radiation like
energy conservation (\ref{FRW4}). However, unlike radiation, the non-local
energy density may be negative.

Nevertheless, if we assume Eq. (\ref{GerU}) then the EFE (\ref{FE4D1}%
-\ref{FE34D}), become%
\begin{align}
3H^{2} &  =\kappa^{2}\rho+\frac{\kappa^{2}}{2\lambda}\rho^{2}+\kappa
^{2}\mathcal{U}+\Lambda,\label{GER1}\\
2\dot{H}+3H^{2} &  =-\kappa^{2}p-\frac{\kappa^{2}}{2\lambda}\left(  2\rho
p+\rho^{2}\right)  -\frac{\kappa^{2}}{3}\mathcal{U}+\Lambda,\label{GER2}\\
\dot{\rho}+3\left(  \rho+p\right)  H &  =0,\qquad\rho=\rho_{0}a^{-3\left(
\gamma+1\right)  },\label{GER3}\\
\mathcal{\dot{U}}+4H\mathcal{U} &  =0,\qquad\mathcal{U=U}_{0}a^{-4}%
.\label{GER4}%
\end{align}

There is no difference between Eqs. (\ref{FRW1}-\ref{FRW4}) and Eqs.
(\ref{GER1}-\ref{GER4}) while the \textquotedblleft
constants\textquotedblright\ ($\kappa^{2},\lambda,\Lambda$) behave as true
\textquotedblleft constants\textquotedblright.

Once the EFE are defined, we would like to study if they are invariant under a
conformal motion, that is, if they admit homothetic or power law solutions. In
particular, we will study if the EFE are invariant under scale
transformations. In fact, this a generalization of Dimensional Analysis (DA)
(see for instance \cite{CT}-\cite{CC05}). To this purpose, we apply the matter
collineations method, that is, we calculate the Lie derivative of the
effective energy momentum tensor along a homothetic vector field
(\cite{Hall}). We would like to point out that there are no self-similar or
power-law solutions for the FE (\ref{FRW1}-\ref{FRW4}), since the same
physical quantity, $\rho,$ has different order of magnitude.

By taking into account the homothetic vector field for the metric
(\ref{FLRW}), that is \cite{HW},
\begin{equation}
\mathcal{H}=t\partial_{t}+\sum_{i=1}^{3}\left(  1-tH\right)  x_{i}%
\partial_{x_{i}},\label{HO}%
\end{equation}
then, we calculate the Lie derivative for each component of the right hand
side of Eq. (\ref{FE4D}), that is:%
\begin{align}
\mathcal{L}_{\mathcal{H}}T_{ij}^{\mathrm{m}} &  =0\qquad\Longleftrightarrow
\qquad\dot{\rho}t+2\rho=0,\qquad\Longleftrightarrow\qquad\rho=\rho_{0}%
t^{-2},\label{Ay1}\\
\mathcal{L}_{\mathcal{H}}S_{ij} &  =0\qquad\Longleftrightarrow\qquad\dot{\rho
}t+\rho=0,\qquad\Longleftrightarrow\qquad\rho=\rho_{0}t^{-1},\label{Ay2}\\
\mathcal{L}_{\mathcal{H}}\mathcal{E}_{ij} &  =0\qquad\Longleftrightarrow
\qquad\mathcal{\dot{U}}t+2\mathcal{U}=0,\qquad\Longleftrightarrow
\qquad\mathcal{U}=\mathcal{U}_{0}t^{-2},\label{Ay3}%
\end{align}
where, it is worth noting that Eqs. (\ref{Ay1} and \ref{Ay2}) are
contradictory, they imply that
\begin{equation}
k_{1}t^{-2}=G\rho_{0}t^{-2}+\frac{G}{2\lambda}\rho^{2}+\frac{6}{\lambda
\kappa^{2}}\mathcal{U}_{0}t^{-2}+\Lambda,
\end{equation}
therefore, there are no self-similar or power-law solutions for this model.
This result is independent of the definition given for $\mathcal{E}_{ij}.$
Note furthermore that if $\Lambda\neq0,$ then the resulting FE cannot admit
this kind of solutions. In order to admit such solutions,, $\Lambda$ must
vary, $\dot{\Lambda}\neq0$. The conclusion is also valid for any homogeneous
self-similar metric (Bianchi models, for example) since their homothetic
vector field always takes the from $\mathcal{H}=t\partial_{t}+...$. (see for
instance \cite{WE}). This kind of models are called spatially homogeneous,
that is, the $G_{6}$ FLRW and the $G_{4}$ models. If the $G_{4}$ has a
subgroup $G_{3}$ which acts simply transitively on the three-dimensional
orbits, we obtain the LRS Bianchi models. Otherwise, we obtain the
Kantowski-Sachs models, and the SS Bianchi models. In addition, although the
model evolves in time, the evolution is fully specified by this symmetry
property, and the FE are purely algebraic when expressed in terms of
expansion-normalized variables.

\section{Brane world models with dynamical brane tension}

In this section, we introduce the idea of a brane world model with variable
tension. We start by considering the FE is 5D, ignoring the contribution of
possible nonstandard model fields in 5D and a symmetric embedding, hence, as
above, we consider the simplest case defined by
\begin{equation}
G_{ab}^{5}=\kappa_{5}^{2}T_{ab}^{5}-\Lambda_{5}g_{ab}^{5}, \label{N5D}%
\end{equation}
where in a first approach we assume that the constants $\kappa_{5}^{2}$ and
$\Lambda_{5}$ are true constants. Following the same procedure, as in the
above section, we define the total energy-momentum tensor on the brane as
$T_{ab}^{5}=\delta\left(  \chi\right)  \left(  T_{ab}^{m}-\lambda
g_{ab}\right)  ,$ and taking into account the Bianchi identities, we impose
that the right hand side of Eq. (\ref{N5D}) has divergence zero, that is
\begin{equation}
\nabla^{b}\left(  T_{ab}^{m}\right)  =\nabla_{a}\lambda, \label{NCE}%
\end{equation}
where obviously there is no matter conservation $\left(  \nabla^{b}\left(
T_{ab}^{m}\right)  \neq0\right)  $, and it is possible to give a
thermodynamical interpretation of this new model as done, for example, by Wong
et al in (\cite{WCH10}). In the case of variable brane tension, the projected
4D EFE on the brane have a similar form to the general case, therefore they
are%
\begin{equation}
G_{ij}=\kappa^{2}T_{ij}^{m}+\kappa_{5}^{4}S_{_{ij}}-\mathcal{E}_{_{ij}%
}\,-\Lambda g_{_{ij}}, \label{N4D}%
\end{equation}
where%
\begin{equation}
\kappa^{2}=\frac{\kappa_{5}^{4}}{6}\lambda,\qquad\Lambda=\frac{1}{2}\left(
\kappa_{5}^{2}\Lambda_{5}+\kappa^{2}\lambda\right)  , \label{defGL}%
\end{equation}
so the assumption of variable brane tension imply that the 4D gravitational
and cosmological \textquotedblleft constant\textquotedblright\ must vary. Note
that from Eq. (\ref{defGL}) we may assume that $\kappa^{2}$ must vary as
$\lambda$ in order to keep $\kappa_{5}^{4}$ constant and therefore $\Lambda$
varies as $\lambda^{2}.$ However, the evolution and conservation equations
change in this framework. In order to deduce the new equations, we impose the
Bianchi identity to the right hand side of Eq. (\ref{N4D}), that is
\begin{equation}
\nabla^{j}\left(  \kappa^{2}T_{ij}^{m}+\kappa_{5}^{4}S_{_{ij}}-\mathcal{E}%
_{_{ij}}\,-\Lambda g_{_{ij}}\right)  =0,
\end{equation}
with the constraint $\nabla^{j}T_{ij}^{m}=\nabla^{j}\lambda.$ Therefore we may
consider
\begin{equation}
\nabla^{j}\mathcal{E}_{_{ij}}=\nabla^{j}\left(  \kappa^{2}T_{ij}^{m}\right)
+\kappa_{5}^{4}\nabla^{j}S_{_{ij}}-\nabla^{j}\left(  \Lambda g_{_{ij}}\right)
, \label{contvc}%
\end{equation}
where $\mathcal{E}_{_{ij}}$ also depends on $\kappa^{2}$ and $\lambda.$ From
Eq. (\ref{defGL}) we get:%
\begin{equation}
\nabla^{j}\kappa^{2}=\kappa^{2}\frac{\nabla^{j}\lambda}{\lambda},\qquad
\nabla^{j}\Lambda=\kappa^{2}\nabla^{j}\lambda. \label{key}%
\end{equation}

Simplifying Eq. (\ref{contvc}) with (\ref{key} and \ref{NCE}) we get%
\begin{equation}
\nabla^{j}\mathcal{E}_{_{ij}}=T_{ij}^{m}\kappa^{2}\frac{\nabla^{j}\lambda
}{\lambda}+\frac{6\kappa^{2}}{\lambda}\nabla^{j}S_{_{ij}},
\end{equation}
and therefore the EFE become%
\begin{align}
G_{ij}  &  =\kappa^{2}T_{ij}^{m}+\kappa_{5}^{4}S_{_{ij}}-\mathcal{E}_{_{ij}%
}\,-\Lambda g_{_{ij}},\label{TVC1}\\
\nabla^{j}T_{ij}^{m}  &  =\nabla_{i}\lambda,\label{TVC2}\\
\nabla^{j}\mathcal{E}_{_{ij}}  &  =T_{ij}^{m}\kappa^{2}\frac{\nabla^{j}%
\lambda}{\lambda}+\frac{6\kappa^{2}}{\lambda}\nabla^{j}S_{_{ij}}. \label{TVC3}%
\end{align}

For a perfect fluid and the flat FLRW metric with $\mathcal{E}_{_{ij}}$
defined by (\ref{DefU}), the EFE (\ref{TVC1}-\ref{TVC3}) yield%
\begin{align}
3H^{2}  &  =\kappa^{2}\rho+\frac{\kappa^{2}}{2\lambda}\rho^{2}+\frac
{6}{\lambda\kappa^{2}}\mathcal{U}+\Lambda,\label{IGA1}\\
2\dot{H}+3H^{2}  &  =-\kappa^{2}p-\frac{\kappa^{2}}{2\lambda}\left(  2\rho
p+\rho^{2}\right)  -\frac{2}{\lambda\kappa^{2}}\mathcal{U}+\Lambda
,\label{IGA2}\\
\dot{\rho}+3\left(  \rho+p\right)  H  &  =-\dot{\lambda},\label{IGA3}\\
\mathcal{\dot{U}}+4H\mathcal{U}-2\mathcal{U}\frac{\dot{\lambda}}{\lambda}  &
=-\frac{5}{6}\rho\kappa^{4}\dot{\lambda}. \label{IGA4}%
\end{align}
Note that if $\mathcal{U}=0,$ or $\rho=0,$ then $\dot{\lambda}=0,$ that is,
the geometry of the bulk must have a non vanishing Weyl tensor or the matter
model of the brane is not the vacuum.

If $\mathcal{E}_{_{ij}}$ is defined by (\ref{GerU}) then the EFE
(\ref{TVC1}-\ref{TVC3}) become
\begin{align}
3H^{2}  &  =\kappa^{2}\rho+\frac{\kappa^{2}}{2\lambda}\rho^{2}+\kappa
^{2}\mathcal{U}+\Lambda,\label{SIRI1}\\
2\dot{H}+3H^{2}  &  =-\kappa^{2}p-\frac{\kappa^{2}}{2\lambda}\left(  2\rho
p+\rho^{2}\right)  -\frac{1}{3}\kappa^{2}\mathcal{U}+\Lambda,\label{SIRI2}\\
\dot{\rho}+3\left(  \rho+p\right)  H  &  =-\dot{\lambda},\label{SIRI3}\\
\mathcal{\dot{U}}+4H\mathcal{U}+\mathcal{U}\frac{\dot{\lambda}}{\lambda}  &
=-5\rho\frac{\dot{\lambda}}{\lambda}. \label{SIRI4}%
\end{align}
noting that the main difference with respect to Eqs. (\ref{IGA1}-\ref{IGA4})
is given by Eq. (\ref{SIRI4}).

\subsection{Matter collineation approach}

As above, we study if these new EFE may admit homothetic solutions. By taking
into account the matter collineation approach we can determine the behavior of
each physical quantity. Hence%
\begin{align}
\mathcal{L}_{\mathcal{H}}\left(  \Lambda g_{ij}\right)   &  =0\qquad
\Longleftrightarrow\qquad\frac{\dot{\Lambda}}{\Lambda}=\frac{\dot{\kappa}^{2}%
}{\kappa^{2}}+\frac{\dot{\lambda}}{\lambda}=-\frac{2}{t},\qquad
\Longleftrightarrow\qquad\Lambda\thickapprox\kappa^{2}\lambda\thickapprox
\Lambda_{0}t^{-2},\label{Z1}\\
\mathcal{L}_{\mathcal{H}}\left(  \kappa^{2}T_{ij}^{\mathrm{m}}\right)   &
=0\qquad\Longleftrightarrow\qquad\frac{\dot{\kappa}^{2}}{\kappa^{2}}%
+\frac{\dot{\rho}}{\rho}=-\frac{2}{t},\qquad\Longleftrightarrow\qquad
\kappa^{2}\rho\thickapprox t^{-2},\label{Z2}\\
\mathcal{L}_{\mathcal{H}}\left(  \kappa_{5}^{4}S_{ij}\right)   &
=0\qquad\Longleftrightarrow\qquad\frac{\dot{\rho}}{\rho}=-\frac{1}{t}%
,\qquad\Longleftrightarrow\qquad\rho\thickapprox t^{-1},\label{Z3}\\
\mathcal{L}_{\mathcal{H}}\left(  \mathcal{E}_{ij}\right)   &  =0\qquad
\Longleftrightarrow\qquad\frac{\mathcal{\dot{U}}}{\mathcal{U}}-\frac
{\dot{\kappa}^{2}}{\kappa^{2}}-\frac{\dot{\lambda}}{\lambda}=-\frac{2}%
{t},\qquad\Longleftrightarrow\qquad\frac{\mathcal{U}}{\lambda\kappa^{2}%
}\thickapprox t^{-2}.\label{Z4}%
\end{align}
where in Eq. (\ref{Z4}) we are considering its definition given by Eq.
(\ref{DefU}).

Therefore, we may conclude that each physical quantity must behave as
\begin{equation}
a\thicksim t^{a_{1}},\qquad\rho\thicksim t^{-1},\qquad\kappa^{2}\thicksim
t^{-1},\qquad\Lambda\thicksim t^{-2},\qquad\lambda\thicksim t^{-1}%
,\qquad\mathcal{U}\thicksim t^{-4},\label{CH2}%
\end{equation}
in order to obtain homothetic or power law solutions. At this point, we would
like to emphasize one important theoretical issue. If we set $\Lambda\thicksim
t^{-2},$ that is, $\Lambda=\Lambda_{0}t^{-2},$ where $\Lambda_{0}$ is a
dimensional constant, this implies that from the definition of $\Lambda,$
$2\Lambda=\left(  \kappa_{5}^{2}\Lambda_{5}+\kappa^{2}\lambda\right)  ,$ that
$\Lambda_{5}$ must be zero, $\Lambda_{5}=0,$ otherwise we have to write
$\Lambda=K+\Lambda_{0}t^{-2},$ $K=const.,$ which is a kinematic self-similar
solution instead of self-similar one (see for instance \cite{CC05}). So for
the self- similar solutions we are assuming that $\Lambda_{5}=0$.

As can be observed, we have obtained $\kappa^{2}\thickapprox\lambda,$ without
any assumption. This result, which is not new in the literature, has been
assumed (or imposed) by several authors (see for instance \cite{NCD1}%
-\cite{NCD2}-\cite{AV} and \cite{BKM}). Note furthermore that,\textbf{ }
$\kappa^{2}\thicksim\lambda\thicksim t^{-1},$ so the gravitational
\textquotedblleft constant\textquotedblright\ is always decreasing as the
brane tension. Therefore, in the framework of the homothetic solutions, the
model always predicts a decreasing gravitational constant.

These relationships keep the 5D constants as true constants. At this point we
have only obtained the order of magnitude for each quantity. Now, we have to
find the exact values of the numerical constants $a_{1},$ $\rho_{0},$
$\kappa_{0}^{2}$, etc, in order to find an exact solution, if it is possible,
since we have six unknowns and only four equations. These results (\ref{CH2})
are dimensionally consistent. By inserting them into the system of equations
(\ref{IGA1}-\ref{IGA4}), we may observe that the system collapses to an
algebraic one, as expected, and whose solution is:%
\begin{align}
a_{1} &  =\frac{\lambda_{0}+\rho_{0}}{3\rho_{0}\left(  \gamma+1\right)
},\quad\kappa_{0}^{2}=\frac{2\lambda_{0}\left[  -\left(  3\gamma+1\right)
\rho_{0}^{2}+\left(  1-3\gamma\right)  \lambda_{0}\rho_{0}+2\lambda_{0}%
^{2}\right]  }{3\rho_{0}\left(  \gamma+1\right)  ^{2}\left[  -\left(
3\gamma+1\right)  \rho_{0}^{2}+\left(  11-3\gamma\right)  \lambda_{0}\rho
_{0}+2\lambda_{0}^{2}\right]  },\nonumber\\
\Lambda_{0} &  =\frac{\lambda_{0}\left[  -5\left(  3\gamma+1\right)  \rho
_{0}^{3}+2\left(  2-9\gamma\right)  \lambda_{0}\rho_{0}^{2}+\left(
11-3\gamma\right)  \lambda_{0}^{2}\rho_{0}+2\lambda_{0}^{2}\right]  }%
{3\rho_{0}\left(  \gamma+1\right)  ^{2}\left[  -\left(  3\gamma+1\right)
\rho_{0}^{2}+\left(  11-3\gamma\right)  \lambda_{0}\rho_{0}+2\lambda_{0}%
^{2}\right]  },\label{RES1}\\
\mathcal{U}_{0} &  =\frac{5\left(  \lambda_{0}+\rho_{0}\right)  \left[
-\left(  3\gamma+1\right)  \rho_{0}^{2}+\left(  1-3\gamma\right)  \lambda
_{0}\rho_{0}+2\lambda_{0}^{2}\right]  \lambda_{0}^{3}}{9\rho_{0}^{2}\left(
\gamma+1\right)  ^{3}\left[  -\left(  3\gamma+1\right)  \rho_{0}^{2}+\left(
11-3\gamma\right)  \lambda_{0}\rho_{0}+2\lambda_{0}^{2}\right]  ^{2}%
}.\nonumber
\end{align}

Since each numerical constant (with dimensions) $a_{1},$ $\rho_{0},$
$\kappa_{0}^{2}$, etc, depends on three parameters ($\lambda_{0},\rho
_{0},\gamma$) then, it is necessary to study them numerically. For example, we
have fixed, $\lambda_{0}=2>0,$ and plotted each parameter in the plane
($\rho_{0},\gamma$) showing that if $\gamma\in(-1,1],$ and $\rho_{0}\in(0,1]$,
then:
\begin{equation}
a_{1}>0,\qquad\kappa_{0}^{2}>0,\qquad\Lambda_{0}>0,\qquad\mathcal{U}_{0}>0,
\end{equation}
which, at least, is physically reasonable. We can even find regions within the
subset $\gamma\in(-1,1],$ and $\rho_{0}\in(0,1]$ where $a_{1}>1$, so the
solution is accelerative. For values of $\gamma<-1$ some of this numerical
constants take negative values. Hence, we may conclude that $\gamma$ can only
take values in the interval $(-1,1].$ This approach has one advantage which is
related to the possibility of determining the integration constants that
appear in the theoretical models. Usually, these constants are determined from
the initial conditions, which are not or poorly known. The present approach
allows an independent determination of the free parameters of the different
cosmological models, leading to the possibility of the direct confrontation of
the theoretical results with observations.

To solve the system of equations (\ref{SIRI1}-\ref{SIRI4}), we need to
determine the behavior of $\mathcal{U},$ since we are taking into account the
definition of $\mathcal{E}_{ij}$ given by Eq. (\ref{GerU}). If we consider
this definition, then we get:
\begin{equation}
\mathcal{L}_{\mathcal{H}}\left(  \mathcal{E}_{ij}\right)  =0\qquad
\Longleftrightarrow\qquad\frac{\mathcal{\dot{U}}}{\mathcal{U}}+\frac
{\dot{\kappa}^{2}}{\kappa^{2}}=-\frac{2}{t},\qquad\Longleftrightarrow
\qquad\kappa^{2}\mathcal{U}\thickapprox t^{-2}.
\end{equation}
which implies that $\mathcal{U}\thicksim t^{-1}.$ Hence, in this case the
quantities behave as:%
\begin{equation}
a\thicksim t^{a_{1}},\qquad\rho\thicksim t^{-1},\qquad\kappa^{2}\thicksim
t^{-1},\qquad\Lambda\thicksim t^{-2},\qquad\lambda\thicksim t^{-1}%
,\qquad\mathcal{U}\thicksim t^{-1}.\label{ANGI}%
\end{equation}
The main and unique difference between (\ref{CH2}) and (\ref{ANGI}) is
\begin{equation}
\mathcal{U}\thicksim t^{-4},\quad\mathcal{U}\thicksim t^{-1}.
\end{equation}
This is an important issue, since they are different physical quantities, with
different dimensional equations and therefore they bring us to obtain
different cosmological implications.

We found the following solution for the system (\ref{SIRI1}-\ref{SIRI4})%
\begin{align}
a_{1} &  =\frac{2\mathcal{U}_{0}+5\rho_{0}}{4\mathcal{U}_{0}},\quad\kappa
_{0}^{2}=\frac{3\left[  75\left(  \gamma+1\right)  ^{2}\rho_{0}^{2}+4\left(
3\gamma+1\right)  \mathcal{U}_{0}\left(  5+\mathcal{U}_{0}\right)  \right]
}{2\mathcal{U}_{0}\left[  45\left(  \gamma+1\right)  ^{2}\rho_{0}^{2}+6\left(
3\gamma+13\right)  \left(  \gamma+1\right)  \mathcal{U}_{0}\rho_{0}+8\left(
3\gamma+1\right)  \mathcal{U}_{0}^{2}\right]  },\nonumber\\
\Lambda_{0} &  =\frac{3\rho_{0}\left[  1125\left(  \gamma+1\right)  ^{2}%
\rho_{0}^{3}+450\left(  3\gamma+5\right)  \left(  \gamma+1\right)
\mathcal{U}_{0}\rho_{0}^{2}+20\left(  90\gamma+27\gamma^{2}+47\right)
\mathcal{U}_{0}^{2}\rho_{0}+8\left(  3\gamma+11\right)  \left(  3\gamma
+1\right)  \mathcal{U}_{0}^{3}\right]  }{16\mathcal{U}_{0}^{2}\left[
45\left(  \gamma+1\right)  ^{2}\rho_{0}^{2}+6\left(  3\gamma+13\right)
\left(  \gamma+1\right)  \mathcal{U}_{0}\rho_{0}+8\left(  3\gamma+1\right)
\mathcal{U}_{0}^{2}\right]  },\label{RES2}\\
\lambda_{0} &  =\frac{15\left(  \gamma+1\right)  \rho_{0}^{2}+2\left(
3\gamma+1\right)  \mathcal{U}_{0}\rho_{0}}{4\mathcal{U}_{0}},\nonumber
\end{align}
where from $\lambda_{0}$ we may obtain $\mathcal{U}_{0}=\mathcal{U}_{0}\left(
\rho_{0},\lambda_{0},\gamma\right)  $
\begin{equation}
\mathcal{U}_{0}=\frac{15\rho_{0}\left(  \gamma+1\right)  }{2\left[  -\left(
3\gamma+1\right)  \rho_{0}+2\lambda_{0}\right]  },
\end{equation}
and to replace this result in the other numerical constants given in
(\ref{RES2}) in order to compare them with (\ref{RES1}). The performed
numerical analysis indicates that in the subset of the plane ($\rho_{0}%
,\gamma$) given by $\left(  \rho_{0}\times\gamma\right)  $ with $\rho_{0}%
\in(0,1]$ and $\gamma\in(-1,1],$%
\begin{equation}
a_{1}>0,\qquad\kappa_{0}^{2}>0,\qquad\Lambda_{0}>0,\qquad\lambda_{0}>0,
\end{equation}
that is, this solution has similar behavior as the above one given by
(\ref{RES1}). The main difference is in the behavior of $a_{1}.$ If
$\mathcal{U}_{0}$ is large enough in the region $\left(  (0,1]\times
(-1,1]\right)  $, then $a_{1}\longrightarrow1/2,$ and therefore the solution
does not accelerate.

\section{Another approach}

We may now reason in a different way. If we consider $\kappa_{5}^{4}%
=6\kappa^{2}/\lambda$, and redo the calculations, in particular the Eq.
(\ref{Z3})\textbf{, }we get again \
\begin{align}
\mathcal{L}_{\mathcal{H}}\left(  \Lambda g_{ij}\right)   &  =0\qquad
\Longleftrightarrow\qquad\dot{\Lambda}t+2\Lambda=0,\qquad\Longleftrightarrow
\qquad\Lambda=\Lambda_{0}t^{-2},\label{J1}\\
\mathcal{L}_{\mathcal{H}}\left(  \kappa^{2}T_{ij}^{\mathrm{m}}\right)   &
=0\qquad\Longleftrightarrow\qquad\frac{\dot{\kappa}^{2}}{\kappa^{2}}%
+\frac{\dot{\rho}}{\rho}=-\frac{2}{t},\qquad\Longleftrightarrow\qquad
\kappa^{2}\rho\thickapprox t^{-2},\label{J2}\\
\mathcal{L}_{\mathcal{H}}\left(  \frac{\kappa^{2}}{\lambda}S_{ij}\right)   &
=0\qquad\Longleftrightarrow\qquad\frac{\dot{\kappa}^{2}}{\kappa^{2}}%
-\frac{\dot{\lambda}}{\lambda}+2\frac{\rho^{\prime}}{\rho}=-\frac{2}{t}%
,\qquad\Longleftrightarrow\qquad\frac{\kappa^{2}}{\lambda}\rho^{2}\thickapprox
t^{-2},\label{JJ3}\\
\mathcal{L}_{\mathcal{H}}\left(  \mathcal{E}_{ij}\right)   &  =0\qquad
\Longleftrightarrow\qquad\frac{\mathcal{\dot{U}}}{\mathcal{U}}-\frac
{\dot{\kappa}^{2}}{\kappa^{2}}-\frac{\dot{\lambda}}{\lambda}=-\frac{2}%
{t},\qquad\Longleftrightarrow\qquad\frac{\mathcal{U}}{\kappa^{2}\lambda
}\thickapprox t^{-2}.\label{J4}%
\end{align}
with $\mathcal{E}_{ij}$ given by Eq. (\ref{DefU}). From (\ref{JJ3}), we get
$\kappa^{2}\rho^{2}/\lambda\thickapprox t^{-2}$, while from (\ref{J2}) we have
obtained $\kappa^{2}\rho\thickapprox t^{-2},$ and therefore we get a new
relationship%
\begin{equation}
\frac{\lambda^{\prime}}{\lambda}=\frac{\rho^{\prime}}{\rho}\Longleftrightarrow
\lambda\thickapprox\rho,\label{J3_2}%
\end{equation}
that is, $\lambda$ and $\rho$ behave in the same way, $\lambda\thickapprox
\rho$. From the definition (\ref{DT}) we observe that these two quantities
have the same dimensional equation, $\left[  \lambda\right]  =\left[
\rho\right]  ,$ and therefore, from this point of view, they must behave in
the same way. If we take into account all these considerations, we may check
that all the EFE are dimensionally consistent (with $\left[  c\right]  =1$),
for example, the Friedman equation
\begin{equation}
\left[  H^{2}\right]  =T^{-2}=\left[  \kappa^{2}\rho\right]  =\left[
\frac{\kappa^{2}}{\lambda}\rho^{2}\right]  =\left[  \frac{\mathcal{U}}%
{\lambda\kappa^{2}}\right]  =\left[  \Lambda\right]  .
\end{equation}

Now, from (\ref{J4}) we obtain $\mathcal{U}\thicksim t^{-4}$, since
$\lambda\thickapprox\rho$ and $\kappa^{2}\rho\thickapprox t^{-2}$ and hence
$\kappa^{2}\lambda\thickapprox t^{-2}.$ Therefore we have the following
relationships%
\begin{equation}
a\thicksim t^{a_{1}},\qquad\rho\thicksim t^{r},\qquad\kappa^{2}\thicksim
t^{g},\qquad\Lambda\thicksim t^{-2},\qquad\lambda\thicksim t^{r}%
,\qquad\mathcal{U}\thicksim t^{-4},\label{CH1}%
\end{equation}
such that $g=-2-r.$ Note that these new results imply that, $\kappa_{5}%
^{2}\neq const.$ If $r=-1,$ then we may recover the solution (\ref{CH2}) as a
particular case. While, if we take into account Eq. (\ref{GerU}), then the
quantities must behave as%
\begin{equation}
a\thicksim t^{a_{1}},\qquad\rho\thicksim t^{r},\qquad\kappa^{2}\thicksim
t^{-2-r},\qquad\Lambda\thicksim t^{-2},\qquad\lambda\thicksim t^{r}%
,\qquad\mathcal{U}\thicksim t^{r}.\label{CH1a}%
\end{equation}

Such solutions (\ref{CH1}-\ref{CH1a}) open the door to having a growing $G$ if
$r<-2,$ but we cannot use them to find solutions from Eqs. (\ref{TVC1}%
-\ref{TVC3}), since they have been deduced under the assumption that
$\kappa_{5}^{2}$ and $\Lambda_{5}$ are constants. Therefore, we need to
propose a more general framework in order to accommodate the possibility of
having a $\kappa_{5}^{2}\neq const.$

For example, if we fix the scale factor, $a\thicksim\exp(a_{1}t),$ then
$\kappa^{2}$ must be growing in order to keep $\kappa^{2}\rho$ const. In the
same way, $\lambda$ must vary as $\rho$ (that is, in a decreasing way) and if
we use (\ref{DefU}), then $\mathcal{U}\sim const,$ while if we use Eq.
(\ref{GerU}) $\mathcal{U}\sim\rho$, that is, it is decreasing. Note that
$\Lambda\thickapprox\kappa^{2}\lambda\sim const,$ hence, it is possible to
find solutions with $\kappa^{2}$ growing and $\lambda$ decreasing such that
$\Lambda\sim const,$ so the hypothesis of a variable $\lambda$ does not
necessary imply variation in all the 4D constants. These results may be
formalized by employing the approach of the matter collineations, but instead
of using a homothetic vector field, we need to use another VF, as for example,
$X=K\partial_{t}-x\partial_{x}-y\partial_{y}-z\partial_{z}\in\mathfrak{X}(M),$
$K\in\mathbb{R},$ since we are looking for other transformations that keep
$H\sim const.$

These results suggest that we may consider $\kappa_{5}^{2}$ and $\Lambda_{5}$
as being variable as in the 4D model. If constants vary in 4D, why shouldn't
they vary in 5D? For example, the Kaluza-Klein models suggest that they must
vary in $n$D (see for instance \cite{Marciano84}, \cite{Chodos80} and
\cite{Brax2003}).

The approach is phenomenological and can be described in the following way. We
consider the gravitational FE in 5D
\begin{equation}
G_{ab}^{5}=\kappa_{5}^{2}T_{ab}^{5}-\Lambda_{5}g_{ab}^{5},
\end{equation}
and impose the condition (\cite{Lau}-\cite{Lau2})%
\begin{equation}
\nabla^{b}(\kappa_{5}^{2}T_{ab}^{5}-\Lambda_{5}g_{ab})=0
\end{equation}
where $T_{ab}^{5}$ is given by Eq. (\ref{DT}). Therefore, the above equation
maybe written as:
\begin{equation}
\nabla^{b}(\kappa_{5}^{2}T_{ab}^{m})=\nabla^{b}(\kappa_{5}^{2}\lambda
g_{ab})+\nabla^{b}(\Lambda_{5}g_{ab}),
\end{equation}
hence%
\begin{equation}
\nabla^{b}(T_{ab}^{m})=\frac{1}{\kappa_{5}^{2}}\left[  \kappa_{5}^{2}%
\nabla_{a}\lambda+\lambda\nabla_{a}\kappa_{5}^{2}+\nabla_{a}\Lambda_{5}%
-T_{ab}^{m}\nabla^{b}\kappa_{5}^{2}\right]  .\label{DEL}%
\end{equation}
We may now assume the condition $\nabla^{b}(T_{ab}^{m})=0,$ in order to avoid
matter creation processes (\cite{Prigo}), so simplifying Eq. (\ref{DEL}) we
get%
\begin{equation}
\nabla_{a}\Lambda_{5}=\left(  T_{ab}^{m}+\lambda g_{ab}\right)  \nabla
^{b}\kappa_{5}^{2}-\kappa_{5}^{2}\nabla_{a}\lambda,
\end{equation}
hence, in 4D the EFE yield%
\begin{equation}
G_{ij}=\kappa^{2}T_{ij}^{m}+\kappa_{5}^{4}S_{_{ij}}-\mathcal{E}_{_{ij}%
}\,-\Lambda g_{_{ij}},
\end{equation}
with%
\begin{equation}
\nabla^{j}\left(  \kappa^{2}T_{ij}^{m}+\kappa_{5}^{4}S_{_{ij}}-\mathcal{E}%
_{_{ij}}\,-\Lambda g_{_{ij}}\right)  =0,
\end{equation}
therefore%
\begin{equation}
\nabla^{j}\left(  \mathcal{E}_{_{ij}}\,\right)  =T_{ij}^{m}\nabla^{j}\left(
\kappa^{2}\right)  +\kappa^{2}\nabla^{j}\left(  T_{ij}^{m}\right)  +S_{_{ij}%
}\nabla^{j}\left(  \kappa_{5}^{4}\right)  +\kappa_{5}^{4}\nabla^{j}\left(
S_{_{ij}}\right)  -\nabla^{j}\left(  \Lambda g_{_{ij}}\right)  ,
\end{equation}
that can be simplified by assuming
\begin{equation}
\kappa^{2}=\frac{\kappa_{5}^{4}}{6}\lambda,\qquad\Lambda=\frac{1}{2}\left(
\kappa_{5}^{2}\Lambda_{5}+\kappa^{2}\lambda\right)  ,\qquad\nabla^{b}%
(T_{ab}^{m})=0.
\end{equation}
Under all these assumptions and simplification, the resulting EFE in 4D
become:%
\begin{align}
G_{ij} &  =\kappa^{2}T_{ij}^{m}+\kappa_{5}^{4}S_{_{ij}}-\mathcal{E}_{_{ij}%
}\,-\Lambda g_{_{ij}},\label{Lena1}\\
\nabla^{j}T_{ij}^{m} &  =0\\
\nabla_{i}\Lambda_{5} &  =\left(  T_{ij}^{m}+\lambda g_{ab}\right)  \nabla
^{j}\left(  \kappa_{5}^{2}\right)  -\left(  \kappa_{5}^{2}\right)  \nabla
_{i}\lambda\\
\nabla^{j}\left(  \mathcal{E}_{_{ij}}\,\right)   &  =T_{ij}^{m}\nabla
^{j}\left(  \kappa^{2}\right)  +S_{_{ij}}\nabla^{j}\left(  \frac{6\kappa^{2}%
}{\lambda}\right)  +\kappa_{5}^{4}\nabla^{j}\left(  S_{_{ij}}\right)
-\nabla^{j}\left(  \Lambda g_{_{ij}}\right)  ,\label{Lena4}%
\end{align}
which can be simplified further if we consider the special case $\Lambda
_{5}=0,$ hence
\begin{align}
G_{ij} &  =\kappa^{2}T_{ij}^{m}+\kappa_{5}^{4}S_{_{ij}}-\mathcal{E}_{_{ij}%
}\,-\Lambda g_{_{ij}},\\
\nabla^{j}T_{ij}^{m} &  =0,\\
\kappa_{5}^{2}\nabla_{i}\lambda &  =\left(  T_{ij}^{m}+\lambda g_{ab}\right)
\nabla^{j}\kappa_{5}^{2},\\
\nabla^{j}\left(  \mathcal{E}_{_{ij}}\,\right)   &  =T_{ij}^{m}\nabla
^{j}\left(  \kappa^{2}\right)  +S_{_{ij}}\nabla^{j}\left(  \frac{6\kappa^{2}%
}{\lambda}\right)  +\frac{6\kappa^{2}}{\lambda}\nabla^{j}\left(  S_{_{ij}%
}\right)  -\nabla_{i}\left(  \kappa^{2}\lambda\right)  .
\end{align}

These equations are more general than (\ref{TVC1}-\ref{TVC3}) and admit as
solution, for example, (\ref{CH1}) or (\ref{CH1a}) and avoid the case
$\nabla^{j}T_{ij}^{m}\neq0,$ which is always difficult to justify from the
observational point of view.

\section{Conclusions}

In the present paper, we have explored whether the brane-world cosmological
models admit self-similar solutions. We have considered the most simple case
(symmetric embedding) and a perfect fluid as matter source, and we have
determined the behavior of the energy density $\rho.$ We have found that this
class of models do not admit self-similar solutions, since the EFE are not
invariant under scale transformations. The result is quite general and valid
for any self-similar metric.

Nevertheless, if we introduce the hypothesis of variable brane tension, we
have shown that these new equations admit scaling symmetries and then they
admit homothetic solutions. We have studied two cases, depending on the
definition tensor $\mathcal{E}_{ij},$ Eqs. (\ref{DefU} or \ref{GerU}%
)\textbf{.} One drawback that is presented of this phenomenological model is
that it depends crucially on having a non vanishing projected Weyl tensor. If
$\mathcal{E}_{ij}=0,$ then the brane tension must be constant, $\dot{\lambda
}=0,$ and therefore $G$ and $\Lambda$ are constant. If they really do vary,
such \textquotedblleft physical\textquotedblright\ property should not depend
on the embedding.

From a cosmological, as well as a physical point of view, the relevance of the
obtained results is twofold. Firstly, it gives the possibility of obtaining
the functional forms of each physical quantity that admits cosmological power
law solutions of the form $a(t)=t^{a_{1}}$, which are relevant for several
phases of the evolution of the Universe. The deceleration parameter
corresponding to this solution is given by $q=(1/a_{1})-1$, and it can
describe accelerating cosmological models for $a_{1}>1$, decelerating
evolution for $a_{1}<1$, and marginally accelerating solutions for $a_{1}=1$.
Hence, one could obtain severe constraints for the allowed functional form of
each physical quantity by requiring the presence in the theory of this power
law form of the scale factor. Generally, for this form of the scale factor,
the geometrical (Ricci scalar), and physical (energy density) parameters have
a simple time dependence, being inversely proportional to the square of the
time Eqs. (\ref{CH2} and \ref{ANGI}). This indicates that the considered
self-similar solutions do present a singular behavior at $t=0$. The effective
gravitational coupling has a power law dependence on time, and, it always
decreases as $G_{eff}\propto t^{-1}$ while the cosmological constant behaves
as $\Lambda\propto t^{-2},$ that is, they are \textquotedblleft
always\textquotedblright\ decreasing. The second implication of the
self-similar approach is related to the possibility of determining the
integration constants that appear in the theoretical models. Usually, these
constants are determined from the initial conditions, which are not, or poorly
known. The present approach allows an independent determination of the free
parameters of the different cosmological models, leading to the possibility of
the direct confrontation of the theoretical results with observations. We have
studied two cases and both could be physically realistic Eqs. (\ref{RES1} and
\ref{RES2})\textbf{.} The self-similar approach for the investigation of the
gravitational theories represents a powerful approach that can give important
information on the mathematical and physical structures of the particular
models. In the present study we have performed such an analysis for the
brane-world cosmological models, with the results of our analysis indicating
that simple power law cosmological models can be obtained in the framework of
this modified gravity theory.

In the above approach, we have worked under the assumption that the
fundamental constants of the theory in 5D must remain constants. The performed
study of invariance of the EFE under scale transformations indicates that it
could be better to consider $\rho\thicksim\lambda$ as a functional
relationship Eqs. (\ref{CH1} and \ref{CH1a}). This result is more general than
the one obtained in (\ref{CH2} or \ref{ANGI})\textbf{ }and we recover\textbf{
}them by fixing $r=-1.$ In this way, we may obtain $G_{eff}\propto t^{-2-r}$
while $\Lambda\propto t^{-2}$ (is always valid). This new result allows the
possibility of getting $G$ as a growing, constant if $r=-2$ (in this limiting
case, $\Lambda$ vanishes) or as a decreasing. However, such working hypothesis
could imply that the fundamental constants in 5D vary as well, and therefore
the EFE (\ref{TVC1}-\ref{TVC3}) are not valid and new ones need to be
derived.  To this purpose, we have outlined a toy model which allows such a
possibility (see Eqs.\textbf{ }(\ref{Lena1}-\ref{Lena4})) and which admits
Eqs. (\ref{CH1} and \ref{CH1a}) as solutions. In conclusion, we think that the
most appropriate framework for considering the variation of the fundamental
constants is the Jordan frame, where the \textquotedblleft
constants\textquotedblright\ are considered as scalar functions (see for
instance \cite{MM01}-\cite{AGS11}).

\begin{acknowledgments}
We would like to thank the anonymous reviewer for comments and suggestions
that helped us to significantly improve our work. J.A.B. is supported by
COMISI\'{O}N NACIONAL DE CIENCIAS Y TECNOLOG\'{I}A through FONDECYT Grant No 11170083.
\end{acknowledgments}

\end{document}